\documentclass[acmtog]{acmart}
\setcitestyle{square}
\usepackage{booktabs} % For formal tables
\usepackage{paralist}
\usepackage{soul}
\definecolor{darkgreen}{RGB}{1, 128, 1}

\newcommand{\figref}[1]{Figure~\ref{#1}}
\newcommand{\tabref}[1]{Table~\ref{#1}}
\newcommand{\equref}[1]{Equation~\ref{#1}}

\newcommand{\secref}[1]{Section~\ref{#1}}

\begin{document}
\title{Optimizing for Aesthetically Pleasing Quadrotor Camera Motion}
% \title{Optimizing for Globally Smooth Quadrotor Camera Motion}
% \title{Understanding and Optimizing for Aesthetically Pleasing \\ Quadrotor Camera Motion}
% \titlenote{Produces the permission block, and
% \subtitle{Extended Abstract}
% \subtitlenote{The full version of the author's guide is available as
%   \texttt{acmart.pdf} document}

% \author{Anonymous Author(s)}
% \authornote{Dr.~Trovato insisted his name be first.}
\author{Christoph Gebhardt}
\affiliation{
  \institution{AIT Lab, ETH Z{\"u}rich}
%  \streetaddress{Universit{\"a}tstrasse 6}
%  \city{Z{\"u}rich}
%  \state{Switzerland}
%  \postcode{8092}
}
\author{Stefan Stev\v{s}i\'{c}}
\affiliation{
  \institution{AIT Lab, ETH Z{\"u}rich}
%  \streetaddress{Universit{\"a}tstrasse 6}
%  \city{Z{\"u}rich}
%  \state{Switzerland}
%  \postcode{8092}
}
\author{Otmar Hilliges}
\affiliation{
  \institution{AIT Lab, ETH Z{\"u}rich}
%  \streetaddress{Universit{\"a}tstrasse 6}
%  \city{Z{\"u}rich}
%  \state{Switzerland}
%  \postcode{8092}
}
\email{firstname.lastname@inf.ethz.ch}

\renewcommand{\shortauthors}{Gebhardt et al.}

\setcopyright{acmlicensed}
\acmJournal{TOG}
\acmYear{2018}\acmVolume{37}\acmNumber{4}\acmArticle{90}\acmMonth{8} \acmDOI{10.1145/3197517.3201390}

%!TEX root = ../proceedings.tex

\begin{abstract}
%Research into generation of quadrotor camera trajectories identified factors that impact the perception of aerial video predominantly from expert feedback. 
In this paper we first contribute a large scale online study ($N\approx400$) to better understand aesthetic perception of aerial video.
The results indicate that it is paramount to optimize smoothness of trajectories across all keyframes. However, for experts timing control remains an essential tool.
Satisfying this dual goal is technically challenging because it requires giving up desirable properties in the optimization formulation. Second,
informed by this study we propose a method that optimizes positional and temporal reference fit \emph{jointly}. This allows to generate globally smooth trajectories, while retaining user control over reference timings. 
The formulation is posed as a variable, infinite horizon, contour-following algorithm.
Finally, a comparative lab study indicates that our optimization scheme outperforms the state-of-the-art in terms of perceived usability and preference of resulting videos. For novices our method produces smoother and better looking results and also experts benefit from generated timings.
\end{abstract}

\keywords{computational design, aerial videography, quadrotor camera tools, trajectory optimization}


\begin{CCSXML}
<ccs2012>
<concept>
<concept_id>10010147.10010178.10010213.10010215</concept_id>
<concept_desc>Computing methodologies~Motion path planning</concept_desc>
<concept_significance>500</concept_significance>
</concept>
<concept>
<concept_id>10010147.10010371</concept_id>
<concept_desc>Computing methodologies~Computer graphics</concept_desc>
<concept_significance>500</concept_significance>
</concept>
<concept>
<concept_id>10010147.10010178.10010213.10010204</concept_id>
<concept_desc>Computing methodologies~Robotic planning</concept_desc>
<concept_significance>500</concept_significance>
</concept>
<concept>
<concept_id>10010520.10010553.10010554.10010558</concept_id>
<concept_desc>Computer systems organization~External interfaces for robotics</concept_desc>
<concept_significance>100</concept_significance>
</concept>
</ccs2012>
\end{CCSXML}

\ccsdesc[500]{Computing methodologies~Computer graphics}
\ccsdesc[500]{Computing methodologies~Motion path planning}
\ccsdesc[500]{Computing methodologies~Robotic planning}
\ccsdesc[100]{Computer systems organization~External interfaces for robotics}

\begin{teaserfigure}
  \includegraphics[width=\textwidth]{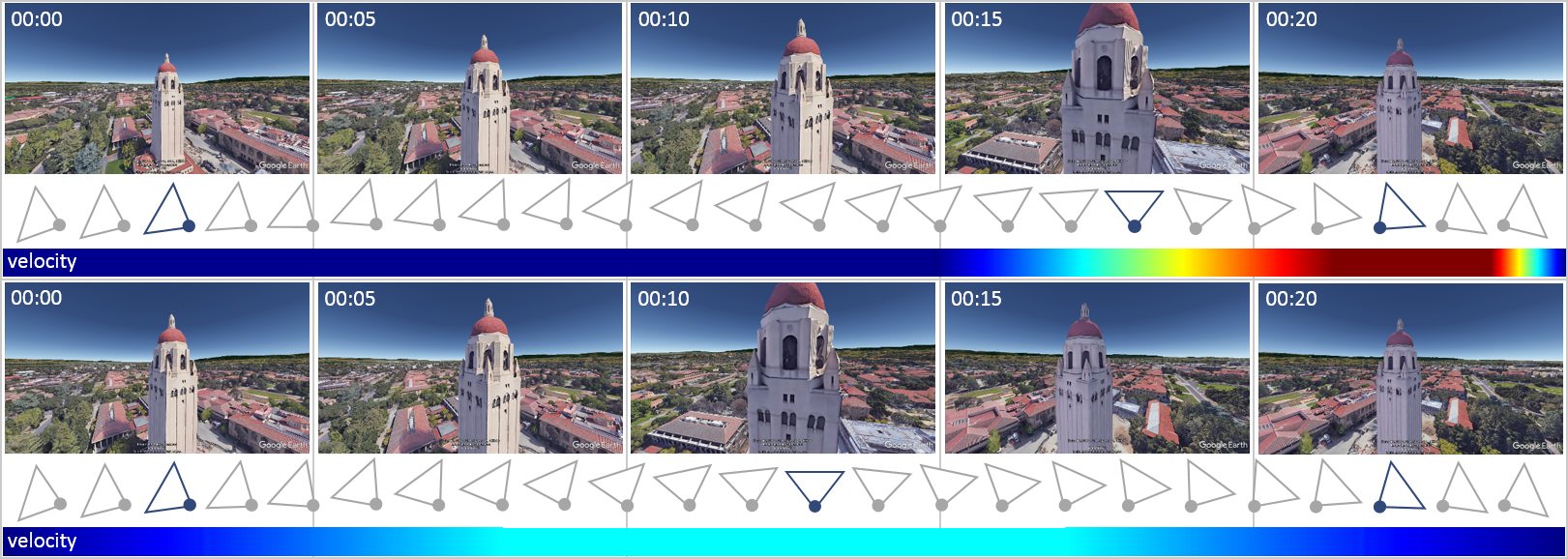}
  \caption{Quadrotor camera tools generate trajectories based on user-specified keyframes in time and space. Reasoning about spatio-temporal distances is hard for users and can lead to visually unappealing results and fluctuating camera velocities. \emph{Top row:} user-specified keyframes (blue) are positioned in time, such that the camera first moves too slow and then needs to accelerate drastically to reach the specified end-point. \emph{Bottom row:} results of our method which automatically positions keyframes (blue) in time such that the camera moves smoothly over the entire trajectory (illustrative example).}
  \label{fig:teaser}
\end{teaserfigure}

\maketitle

%!TEX root = ../submission.tex

\section{Introduction}
Camera quadrotors have become a mainstream technology but fine-grained control of such camera drones for aerial videography is a high-dimensional and hence difficult task.
In response several tools have been proposed to plan quadrotor shots by defining keyframes in virtual environments \cite{Gebhardt:2017:WYFIWYG,Gebhardt:2016:Airways,Joubert2015,roberts:2016}.
This input is then used in an optimization algorithm to automatically generate quadrotor and camera trajectories. %, respecting its physical limits.
Intuitively, smooth camera motion is an obvious factor impacting the visual quality of a shot. This intuition alongside expert-feedback \cite{Joubert2015} and literature on (aerial) cinematography \cite{arijon1976grammar,Audronis:2014,Hennessy:2015} forms the basis for most existing quadrotor tools. These take a spline representation, connecting user specified keyframes, and optimize higher derivatives of these splines, such as jerk.

Balasubramanian et al. \shortcite{Balasubramanian2015} define global smoothness as ``a quality related to the continuity or non-intermittency of a movement, independent of its amplitude and duration''.
However, because keyframe timings are kept fixed in current quadrotor camera optimization schemes \cite{Gebhardt:2016:Airways,Joubert2015}, or close to the user input \cite{roberts:2016}, smooth motion can only be generated subject to these hard-constraints. This can cause strong variation of camera velocities across different trajectory segments and result in visually unpleasant videos. %In the following we detail the problem with an example.

Consider popular fly-by-shots, such as the one illustrated in \figref{fig:teaser}, where an object is filmed first from one direction and then gradually the camera yaws around it's own z-axis by $180^{\circ}$ as the quadrotor flies past the object until it is filmed from the opposing direction. To achieve visually pleasing footage both the quadrotor motion and the camera's angular velocity need to be smooth. Users generally struggle with this or similar problems in which they place the keyframes in the correct \emph{spatial} location but too close (or too far) to each other \emph{temporally} (see \figref{fig:teaser}, top and [\href{https://www.youtube.com/embed/tfcnLkJpRtc?start=42&end=65}{video}]). This is indeed a difficult task because keyframes are specified in 5D (3D position and camera pitch and yaw) and imagining the resulting translational and rotational velocities is cognitively demanding.
%Because existing optimization methods treat user keyframes as hard constraint in time they often cannot produce smooth trajectories in all dimensions.

Although existing work provides UI tools (i.e. progress curves, timelines) to cope with this problem, it has been shown that users, especially novices, struggle to create smooth camera motion over a sequence of keyframes \cite{Gebhardt:2017:WYFIWYG}.
While optimizing for global smoothness may address this issue for novices, an interesting tension arises when looking at experienced users. Experts explicitly time the visual progression of a shot in order to achieve desired compositional effects \cite{Joubert2015} (e.g. ease-in, ease-out behavior).
%Not surprisingly, videos with expert provided timings were rated more favorable in our perceptual online study.
Our first contribution is a large online study ($N=424$), highlighting this issue, where non-expert designed videos were rated more favorable when optimized for global smoothness while expert-designed videos were perceived as more pleasing with hard-constrained timings.  
To the best of our knowledge, this is the first study that provides empirical evidence for global smoothness indeed being important for the perception of aerial videography.
%in the technical sense 

Embracing this dichotomy (of smoothness versus timing control), our second contribution is a trajectory optimization method that takes smoothness as primary objective and can re-distribute robot positions and camera angles in space-time. 
We propose the first algorithm in the area of quadrotor videography that treats keyframe timings and positions, and reference velocities as \emph{soft-constraints}. This extends the state-of-the-art in that it allows users to trade off path-following fidelity with temporal fidelity.
Such a formulation poses significant technical difficulties. Prior methods incorporate keyframe timings as \emph{hard-constraints}, yielding a quadratic and hence convex optimization formulation (depending on the dynamical model), allowing for efficient implementation.
In contrast, we formulated the quadrotor camera trajectory generation problem as a variable, infinite horizon, contour-following algorithm applicable to linear and non-linear quadrotor models. Our formulation has to discretize the model at each solver iteration according to the optimized trajectory end time.
Although this formulation is no-longer convex, it is formulated as well-behaved non-convex problem and our implementation runs at interactive rates.

Finally, we show the benefit of our method compared to the state-of-the-art in a lab study in which we compare different variants of our method with \cite{Gebhardt:2016:Airways}. It can be shown that our method positively effects the usability of quadrotor camera tools and improves the visual quality of video shots for experts and non-experts. Both benefit from using an optimized timing initially, fine-tuning it according to their intention. In addition, the user study revealed that timing control does not need to be precise but is rather used to control camera velocity in order to create a certain compositional effect.

\section{Related Work}
% \subsection{Robotic Behavior Design}
% Automating the design of robotic systems based on high-level functional specifications is a long-standing goal in graphics and HCI. Focusing on robot behavior only, tangible UIs \cite{Zhao:2009:MCP}, and sketch based interfaces to program robotic systems \cite{Liu:2011:RMS,Sakamoto:2009} have been introduced. Recently, several works propose gesture-based interaction with drones \cite{Cauchard:2015,E:2017}.
%and study the perception of such robots by users

\paragraph{Camera Control in Virtual Environments:}
Camera placement \cite{Lino:2012}, path planning~\cite{CAV:CAV398,Li2008} and automated cinematography~\cite{Lino:2011:Director} have been studied extensively in virtual environments (VE), for a survey see~\cite{Christie:2008}. These works share our goal of assisting users in the creation of camera motion (e.g., \cite{Drucker94intelligentcamera,Lino:2015hn}).
%(e.g., ~\cite{Drucker94intelligentcamera,Lino:2015hn,Lino:2011:Director}).
Nevertheless, it is important to consider that VEs are not limited by real-world physics and robot constraints, hence may yield trajectories that can not be flown by a quadrotor.

\paragraph{Character Animation:}
In character animation, a variety of methods exist which are capable of trading-off positional and temporal reference fit to optimize for smoother character motion. In \cite{Liu2006}, the authors specify constraints in warped time and then optimize the mapping between warped and actual time according to their objective function. 
For an original motion, \cite{mccann2006} find the convex hull of all physically valid motions attainable via re-timing. Plausible new motions are then found by performing gradient descent and penalizing distance between possible solutions and the feasible hull. 
Like \cite{Liu2006}, our formulation is based on a time-free parameterization of a reference path. In contrast to the character animation methods, we adjust timings by optimizing the progress of the quadrotor camera on the reference according to a objective favoring smoothness. Unlike \cite{mccann2006} our formulation does not require nested optimization.

\paragraph{Trajectory Generation:}
Trajectory generation for dynamical systems is a well studied problem in
computer graphics \cite{Geijtenbeek:2012} and robotics \cite{Betts:2009}. Approaches that encode the system dynamics as equality constraints to solve for the control inputs along a motion trajectory are referred to as spacetime constraints in graphics \cite{Witkin:1988} and direct collocation in robotics \cite{Betts:2009}. Used out-of-the box such approaches can lead to slow convergence time especially with long time horizons (cf. \cite{roberts:2016}). 

With the commoditization of quadrotors, the generation of drone trajectories shifted into the focus of research. Exploiting the differential flatness of quadrotors in the output
space, \cite{Mellinger:2011} generated physically feasible minimal snap trajectories. Several methods exist for the generation of trajectories for aggressive quadrotor flight \cite{Mellinger2012, Bry:2015}. Traditionally, these methods convert a sequence of input positions into a time-dependent reference and based on a dynamical model generate a trajectory which follows this reference. For \cite{Mellinger:2011, Bry:2015}, time optimization is done in a cascaded manner where an approximated gradient descent for keyframe timings is calculated based on the original optimization problem. 
These formulations suffer from very long runtimes as the original problem needs to be called once for each keyframe to calculate the gradient approximation.  
In contrast, our method optimizes keyframe timings and trajectory jointly reducing optimization runtime and allowing to trade-off temporal and positional fit.
In \cite{Mellinger2012}, sequentially composed controllers are used to optimize the timing of a trajectory such that physical limits are not violated given desired feed-forward terms. Our work does not only ensure physical feasibility but is also capable of generating trajectories with different dynamics (smooth and more aggressive) for the same spatial input. 

% \begin{figure*}[tbh]
% 	\centering
% 	\includegraphics[width=1.0\linewidth]{figures/problem}
% 	\caption{Schematic impact of hard-constraint keyframes. (Top:) Temporally close keyframes (blue) can lead to excessive angular velocities in the generated trajectory (grey), whereas our perceptual study shows that users prefer trajectories with globally smooth velocities (bottom).}
% 	\label{fig:problem}
% \end{figure*}

\paragraph{Computational Support of Aerial Videography:}
A number of tools for the planning of aerial videography exists. Commercially available applications and consumer-grade drones often place waypoints on a 2D map~\cite{apm:2015,dji:2015:groundstation,litchi:2016} or allow to interactively control the quadrotor's camera as it tracks a pre-determined path \cite{3dr:2015:solo}.
These tools generally do not provide means to ensure feasibility of the resulting plans and do not
consider aesthetic or usability objectives in the video composition task.
The planning of physically feasible quadrotor camera trajectories has recently received a lot of attention.
Such tools allow for planning of aerial shots in 3D virtual environments \cite{Gebhardt:2017:WYFIWYG,Joubert2015,Gebhardt:2016:Airways,roberts:2016} and employ optimization to ensure that both aesthetic objectives and robot modeling constraints are considered.

In \cite{Joubert2015} and \cite{Gebhardt:2016:Airways}, users specify keyframes in time and space. These are incorporated as hard-constraints into an objective function. Solving for the trajectory only optimizes camera dynamics and positions. This causes the generation of locally smooth camera motion (between keyframes) but can lead to varying velocities across keyframes. Joubert et al. \shortcite{Joubert2015} detect violations of the robot model constraints. However, correcting these violations is offloaded to the user.
In contrast, by generating timings or incorporating them as soft-constraints our optimization returns the closest feasible fit of the user-specified inputs, subject to our robot model, and generates globally smooth quadrotor camera trajectories.
\cite{Gebhardt:2017:WYFIWYG} re-optimizes keyframe timings in a cascaded optimization scheme. Here an approximated gradient on the keyframe times produced by the optimization formulation of \cite{Gebhardt:2016:Airways} is calculated and used to improve visual smoothness. However, this approach is relatively slow and the paper reports that users therefore did not make significant use of it in the evaluation. In contrast, our method runs at interactive rates optimizing trajectories of different duration within seconds (avg. $2.4~s$).
Roberts and Hanrahan \shortcite{roberts:2016} take physically infeasible trajectories and compute the closest possible feasible trajectory by re-timing the trajectories subject to a non-linear quadrotor model. In contrast, we prevent trajectories from becoming infeasible at optimization time. Although the method of \cite{roberts:2016} theoretically can be used to adjust timings based on a jerk minimization objective, our method can also trade-off the positional fit of a reference path in order to achieve even smoother motion.

Recently, several works address the generation of quadrotor camera trajectories in real-time to record dynamic scenes. \cite{galvane2016automated, joubert2016towards} plan camera motion in a lower dimensional subspace to attain real-time performance. Using a Model Predictive Controller (MPC), \cite{Naegeli2017} optimizes cinematographic constraints, such as visibility and position on the screen, subject to robot constraints for a single quadrotor. \cite{Nageli:2017:SIGGRAPH} extends this work for multiple drones and allows actor-driven tracking on a geometric path. 
Focusing on dynamic scenes, this work does not cover the global planning aspects of aerial videography.

\paragraph{Online Path Planning:}
Approaches that address trajectory optimization and path following have been proposed in the control theory literature. They allow for optimal reference following given real world influences. Methods like MPC \cite{Faulwasser:2009} optimize the reference path and the actuator inputs simultaneously based on the system state. MPC has been successfully use for the real-time generation of quadrotor trajectories \cite{Mueller2013}.
Nevertheless, \cite{AGUIAR2008} show that the tracking error for following timed-trajectories can be larger than for following a geometric path only. Motivated by this observation Model Predictive Contouring Control (MPCC) \cite{Lam2013} has been proposed to follow a time-free reference, optimizing system control inputs for time-optimal progress. MPCC approaches have been successfully applied in industrial contouring \cite{Lam2013} and RC racing \cite{Liniger2014}. Recently, \cite{Nageli:2017:SIGGRAPH} extended the MPCC-framework to allow for real-time path following in 3D space with quadrotors. We propose a trajectory generation method that is conceptually related to MPCC formulations in that it optimizes timings for a quadrotor camera trajectory based on a time-optimal path-following objective. Our formulation treats keyframes, user specified reference timings and velocities as well as smoothness across the entire trajectory jointly in a soft-constrained formulation and allows users to produce aesthetically more pleasing videos.

%!TEX root = ../submission.tex

\section{Method}
We propose a new method to generate \emph{globally smooth} quadrotor camera trajectories. Our aim is to allow even novice users to design complex shots without having to explicitly reason about 5D spatio-temporal distances. Our central hypothesis is that smoothness across the entire trajectory matters and hence is the main objective of our optimization formulation.  We first introduce the model of the system dynamics in \secref{sec:dynamic_model} and discuss our optimization formulation in \secref{sec:variable_horizon}-\ref{sec:optimization_problem}. See Appendix \ref{sec:notation} for a table of notations.
%We do not include a notation section and refer to .
% To validate this hypothesis we ran a large scale perceptual study ($N=424$) which confirmed this hypothesis. However, in agreement with prior work it became clear that experts can and want to control timings explicitly.
% To allow for this fine-grained control we extend our method to take reference timings into consideration.

% Here we \ref{sec:dynamic_model} introduce the model of the system dynamics, \ref{sec:dynamic_model} discuss our optimization formulation and \ref{sec:dynamic_model} its implementation.
% We then
% \begin{inparaenum}[i)]
% 	\item detail the perceptual online study and
% 	\item the extensions to our method informed by insights from that study.
% \end{inparaenum}
% Finally, we discuss a comparative lab study to evaluate the usability of a simple aerial videography planning tool developed on top of the proposed algorithm.

\subsection{Dynamical Model}
\label{sec:dynamic_model}
We use the approximated quadrotor camera model of \cite{Gebhardt:2016:Airways}. This discrete first-order dynamical system is incorporated as equality constraint into our optimization problem:
\begin{align}
\begin{split}
	\label{eq:quad_dynamical_system_discrete}
	\mathbf{x}_{i+1} = A \mathbf{x}_{i} + B \mathbf{u}_{i} + g , \ \mathbf{u}_{min} \leq \mathbf{u}_{i} \leq \mathbf{u}_{max} , \\
    \mathbf{x}_{i} = [\mathbf{r}, \psi_q, \psi_g, \phi_g, \dot{\mathbf{r}}, \dot{\psi_q}, \dot{\psi_g}, \dot{\phi_g}, \ddot{\mathbf{r}}, \ddot{\psi_q}, \ddot{\psi_g}, \ddot{\phi_g}, \dddot{\mathbf{r}}, \dddot{\psi_q}, \dddot{\psi_g}, \dddot{\phi_g}]^{T} , \\
    \mathbf{u}_{i} = [\mathbf{F}, M_{\psi_q}, M_{\psi_g}, M_{\phi_g}]^{T} ,
\end{split}
\end{align}
where $\mathbf{x}_{i} \in \mathbb{R}^{24}$ are the quadrotor camera states and $\mathbf{u}_{i} \in \mathbb{R}^{6}$
are the inputs to the system at stage $i$. Furthermore, $\mathbf{r} \in \mathbb{R}^{3}$ is the position of the quadrotor, $\psi_q$ is the quadrotor's yaw angle and $\psi_g$ and $\phi_g$ are the yaw and pitch angles of the camera gimbal. The matrix
$A \in \mathbb{R}^{24x24}$ propagates
the state $\mathbf{x}$ forward, the matrix
$B \in \mathbb{R}^{24x6}$ defines the effect of the
input $\mathbf{u}$ on the state and the vector
$g \in \mathbb{R}^{24}$ that of gravity for one time-step.
$\mathbf{F}$ is the the force acting on the quadrotor, $M_{\psi_q}$ is the torque along its z-axis and $M_{\psi_g}$, $M_{\phi_g}$ are torques acting on pitch and yaw of the gimbal.

Please note that our formulation is agnostic to the dynamical model of the quadrotor. We verified this by incorporating the non-linear model of \cite{Nageli:2017:SIGGRAPH}. Qualitatively this does not impact results, yet the computational cost increases (see \figref{fig:tech_eval}).

\subsection{Variable Horizon}
\label{sec:variable_horizon}
In space-time optimization, the horizon length is defined by dividing the timing of the last keyframe by the discretization step $\Delta t$.
However, one key idea in our formulation is that we treat trajectories, at least initially, as time free. In particular, our method does not take timed keyframes as input and therefore traditional approaches to determining the horizon length are not applicable.

Taking inspiration from MPC literature \cite{Michalska1993}, we make the length of the horizon an optimization variable itself by adding the trajectory end time $T$ into the state space of our model ($\mathbf{x} = [\mathbf{x}, T]^{T} \in \mathbb{R}^{25}$ with $\frac{\delta T}{\delta t}=0$).
%For this purpose we use the following affine transformation of time $t = \tau \ T$ ($\tau \in [0,1]$), where $T$ is the trajectory end time.
%We then add $T$ into the state space of our model with $\frac{\delta T}{\delta t}=0$, allowing us to optimize the end time of a trajectory.
This has implications for the dynamical model. At each iteration of the solver we adjust the discretization step $\Delta t = \frac{T}{N}$. Here $N$ is the number of stages in the horizon spanning the entire trajectory. The forward propagation matrices $A$ and $B$ are also recalculated based on the current $\Delta t$.

\subsection{Reference Tracking Metric}
We require a time-free parameterization of the reference to optimize the timing of keyframes.
We use a chord length parameterized, piecewise cubic polynomial spline in hermite form (PCHIP) to interpolate the user-defined keyframes \cite{Fritsch1980}. The resulting chord length parameter $\theta$ describes progress on the spatial reference path defined as $\mathbf{f}_d(\theta) = [\mathbf{r}_d(\theta), \psi_d(\theta), \phi_d(\theta)] \in \mathbb{R}^{5}$. 
To prevent sudden changes of the progress parameter, we add $\theta$ into our model and formulate its dynamics with the following linear discrete system equation:
\begin{equation}
	\label{eq:theta_dynamical_system_discrete}
	\mathbf{\Theta}_{i+1} = C \mathbf{\Theta}_{i} + D v_{i}, \ 0 \leq v_{i} \leq v_{max} ,
\end{equation}

where $\mathbf{\Theta}_{i} = [\theta_{i}, \dot{\theta}_{i}]$ is the state and $v_i$ is the input of $\theta$ at step $i$ and $C \in \mathbb{R}^{2x2}$, $D \in \mathbb{R}^{2x1}$ are the discrete system matrices.
Intuitively, $v_i$ approximates the quadrotor's acceleration as $\theta$ is an approximation of the trajectory length.

\begin{figure}[tbh]
	\centering
	\includegraphics[width=0.9\linewidth]{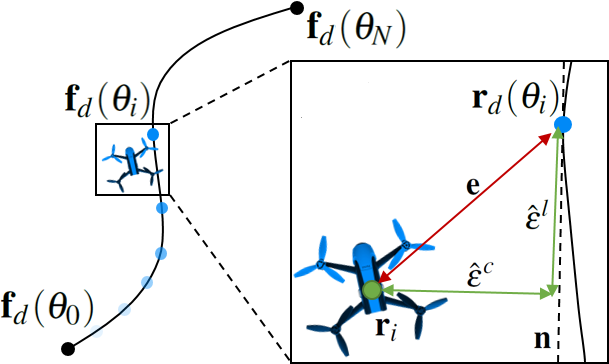}
	\caption{The position (x,y,z) and orientation (yaw, pitch) over time are a function of path progress $\mathbf{f}_d(\theta)$. Inset: to advance along the path we optimize for smooth progress $\dot{\theta}$ via minimization of lag $\hat{\epsilon}^l$ and contour $\hat{\epsilon}^c$ error.}
	\label{fig:lagcontourerror}
\end{figure}

With this extension of the dynamic model in place, we now formulate an objective to minimize the error between the desired quadrotor position $\mathbf{r}_d(\theta)$ and the current quadrotor position $\mathbf{r}$. With respect to the time optimization we want the quadrotor to follow $\mathbf{r}_d(\theta_i)$ as closely as possible in time (no lag) but allow deviations from its contour for smoother motion. This distinction is not possible when minimizing the 2-norm distance to the reference point.
For this reason, we differentiate between a lag $\epsilon^l$ and a contour error $\epsilon^c$ similar to MPCC approaches (e.g., \cite{Lam2013}). We approximate the true error from the spline by using the 3D-space approximation of lag $\hat{\epsilon}^l$ and a contour error $\hat{\epsilon}^c$ of \cite{Nageli:2017:SIGGRAPH} (see \figref{fig:lagcontourerror}, inset).
The approximated lag error is defined as,
\begin{equation}
	\label{eq:lag_error}
	\hat{\epsilon}^l(\theta,\mathbf{r}_i) =  \mathbf{e}^T\mathbf{n}
\end{equation}
where $\mathbf{e} = \mathbf{r}_d(\theta) - \mathbf{r}_i$ is the relative vector between desired and actual positions and $\mathbf{n} = \frac{\dot{\mathbf{r}}_d(\theta)}{||\dot{\mathbf{r}}_d(\theta)||}$ is the normalized tangent vector of $\mathbf{r}_d(\theta)$ at $\theta$. The resulting contour error approximation is given by:
\begin{equation}
	\label{eq:contour_error}
	\hat{\epsilon}^c(\theta, \mathbf{r}_i) =  ||\mathbf{e} - \hat{\epsilon}^l(\theta,\mathbf{r}_i)\mathbf{n}||.
\end{equation}
Both error terms are then inserted into the cost term,
\begin{align}
	\label{eq:position_cost}
    c^{p}(\theta, \mathbf{r}_i)  = \begin{bmatrix}
           \hat{\epsilon}^l(\theta) \\
           \hat{\epsilon}^c(\theta)
         \end{bmatrix}^T
         Q\begin{bmatrix}
           \hat{\epsilon}^l(\theta) \\
           \hat{\epsilon}^c(\theta)
         \end{bmatrix},
\end{align}
where $Q$ is a diagonal positive definite weight matrix. Minimizing $c^{p}$ will move the quadrotor along the user defined spatial reference.

Our experiments have shown that distinguishing between lag and contour error is important for the temporal aspects of the optimization. Trajectories generated by minimizing $||\mathbf{e}||^2$, depending on the weighting of the term, either lag behind the temporal reference or cannot trade-off positional fit for smoother motion. With appropriate weights for lag and contour error this behavior is avoided.

To give users fine grained control over the target framing we follow user-specified viewing angles in an analogous fashion. To attain the camera yaw and pitch we minimize the 2-norm discrepancy between desired and actual orientation of the quadrotor and camera gimbal. Given by the following cost terms:
\begin{align}
	\label{eq:yaw_error}
	c^{\psi}(\theta, \psi_{q,i}, \psi_{g,i})  = || \psi_d(\theta) - (\psi_{q,i} + \psi_{g,i}) || ^ {2} \\
	c^{\phi}(\theta, \phi_{g,i}) = || \phi_d(\theta) - \phi_{g,i} || ^ {2} ,\label{eq:pitch_error}
\end{align}
where $\psi_{q,i}$, $\psi_{g,i}$, $\phi_{g,i}$ are the current yaw and pitch angles. Furthermore, we preprocess every keyframe by adding a multiple of $2\pi$ to yaw and pitch such that the absolute distance to the respective angle of the previous keyframe has the smallest value.

\subsection{Smooth Progress}
For the camera to smoothly follow the path, we need to ensure that $\theta$ progresses. By specifying an initial $\theta_0$ and demanding $\theta$ to reach the end of the trajectory in the terminal state $\theta_N$, the progress of $\theta$ can be forced with an implicit cost term. We simply penalize the trajectory end time by minimizing the state space variable $T$,
\begin{equation}
	\label{eq:end_time_cost}
	c^{end}(T) = T.
\end{equation}
Minimizing the end-time can be interpreted as optimizing trajectories to be as short as possible temporally (while respecting smoothness and limits of the robot model). This forces $\theta$ to make progress such that the terminal state $\theta_N$ is reached within time $T$\footnote{This also prevents solutions of infinitely long trajectories in time where adding steps with $\mathbf{u}_i\approx 0$ is free wrt. to Eq. \eqref{eq:jerk_cost}).}.

To ensure that the generated motion for the quadrotor is smooth, we introduce a cost term on the model's jerk states,
\begin{equation}
	\label{eq:jerk_cost}
	c^j(\dddot{\mathbf{r}}, \dddot{\psi_q}, \dddot{\psi_g}, \dddot{\phi_g}) =  ||\mathbf{j}_i||^2,
\end{equation}
where $\mathbf{j}_{i} = [\dddot{\mathbf{r}}, \dddot{\psi_q}, \dddot{\psi_g}, \dddot{\phi_g}]^{T}$ is jerk and angular jerk. We minimize jerk since it provides a commonly used metric to quantify smoothness \cite{Hogan1984} and is known to be a decisive factor for the aesthetic perception of motion \cite{Bronner2015}.
This cost term again implicitly effects $\theta$ by only allowing it to progress such that the quadrotor motion following the reference path $\mathbf{f}_d(\theta)$ is smooth according to \eqref{eq:jerk_cost}. This is illustrated in \figref{fig:lagcontourerror}, left. The blue dot ($\theta_i$) progresses on the reference path such that the generated motion of the quadrotor following $\mathbf{f}_d(\theta_i)$ is smooth.

To still be able to specify the temporal length of a video shot with this formulation, we define the cost term,
\begin{equation}
	\label{eq:set_end_time}
	c^{len}(N, \Delta t) =  ||t_{len} - T||^2,
\end{equation}
where we minimize the 2-norm discrepancy between the trajectory end time $T$ and a user-specified video length $t_{len}$. In case a trajectory is optimized for Eq. \eqref{eq:set_end_time}, the weight for Eq. \eqref{eq:end_time_cost} is set to zero.

\subsection{Optimization Problem}
\label{sec:optimization_problem}
We construct our overall objective function by linearly combining the cost terms from Eq. \eqref{eq:position_cost}, \eqref{eq:yaw_error}, \eqref{eq:pitch_error}, \eqref{eq:end_time_cost}, \eqref{eq:jerk_cost}, \eqref{eq:set_end_time} and a 2-norm minimization of $v$. The final cost is:
\begin{flalign}
	\label{eq:objective_function}
	&J_i = w_p c^{p}(\theta_i, \mathbf{r}_i) + w_{\psi} c^{\psi}(\theta_i, \psi_{q,i}, \psi_{g,i}) + w_{\phi} c^{\phi}(\theta_i, \phi_{g,i}) \\
    &+ w_j c^j(\dddot{\mathbf{r}}, \dddot{\psi_q}, \dddot{\psi_g}, \dddot{\phi_g})
    + w_{end} c^{end}(T) + w_{len} c^{len}(N, \Delta t) + w_v ||v||^2  , \nonumber
\end{flalign}
where the scalar weight parameters $w_p, w_{\psi}, w_{\phi}, w_j, w_{end}, w_{len}, w_v > 0$ are adjusted for a good trade-off between positional fit and smoothness. 
The final optimization problem is then:
\begin{align}
	\label{eq:optimization_problem}
	\underset{\mathbf{x},\mathbf{u},\mathbf{\Theta},v}{\text{minimize}} \ & \sum_{i=0}^{N} J_i &  \\
	\text{ subject to } & \mathbf{x}_0 = \mathbf{k}_0 & \text{(initial state)} \nonumber \\
	& \mathbf{\Theta}_0 = \mathbf{0} & \text{(initial progress)} \nonumber \\
    & \mathbf{\Theta}_N = \mathbf{L} & \text{(terminal progress)} \nonumber \\
    & \mathbf{x}_{i+1} = A \mathbf{x}_{i} + B \mathbf{u}_{i} + g & \text{(dynamical model)} \nonumber \\
    & \mathbf{\Theta}_{i+1} = C \mathbf{\Theta}_{i} + D v_{i} & \text{(progress model)} \nonumber \\
%     & 0 \leq \theta_{i} \leq L & \text{(progress bounds)} \nonumber \\
    & \mathbf{x}_{min} \leq \mathbf{x}_{i} \leq \mathbf{x}_{max}, & \text{(state bounds)} \nonumber \\
    & \mathbf{u}_{min} \leq \mathbf{u}_{i} \leq \mathbf{u}_{max}, & \text{(input limits)} \nonumber \\
    & \mathbf{0} \leq \mathbf{\Theta}_{i} \leq \mathbf{\Theta}_{max} & \text{(progress bounds)} \nonumber \\
    & \mathbf{0} \leq v_{i} \leq v_{max} & \text{(progress input limits)} \nonumber ,
\end{align}
where $J_i$ is quadratic in $\mathbf{x}$, $\mathbf{u}$, $v$ and linear in $\mathbf{\Theta}$.
When flying a generated trajectory we follow the optimized positional trajectory $\mathbf{r}$ with a standard LQR-controller and use velocity and accelerations states of $\mathbf{x}$ as feed-forward terms.

\begin{figure}[b]
	\centering
	\includegraphics[width=1.0\linewidth]{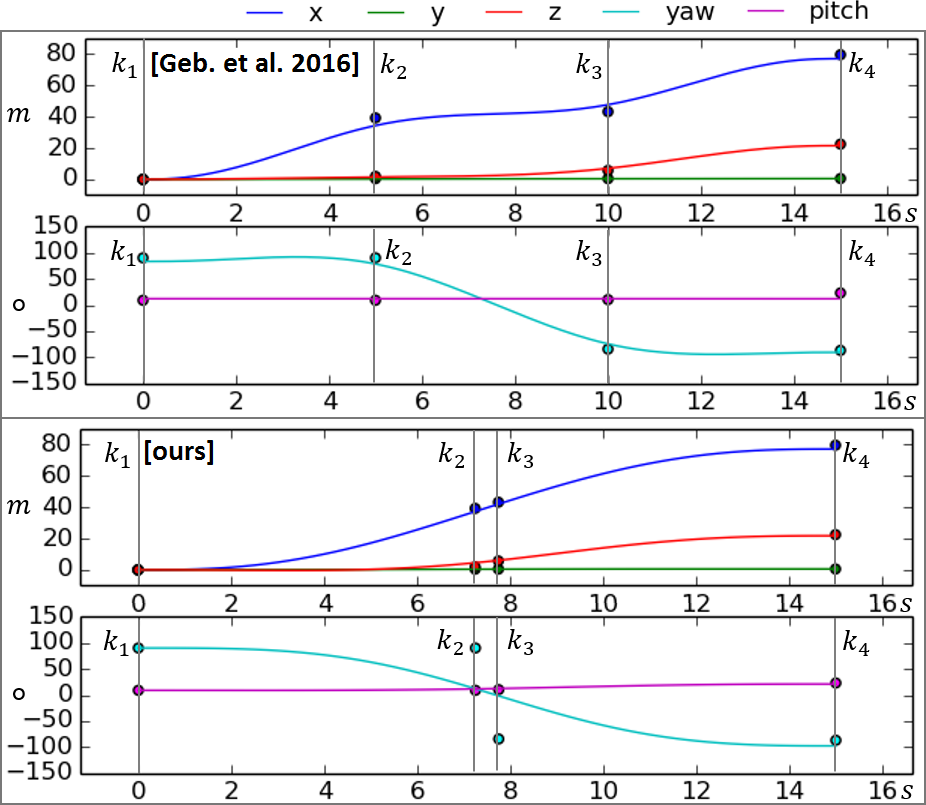}
	\caption{Position (x,y,z in $m$) and orientation (yaw, pitch in $\circ$) over time (in $s$) for the same user-specified keyframes ($k_1$-$k_4$) for \protect\cite{Gebhardt:2016:Airways} (top) and our method (bottom).}
	\label{fig:posPlot}
\end{figure}

\begin{figure*}[tbh]
	\centering
	\includegraphics[width=1.0\linewidth]{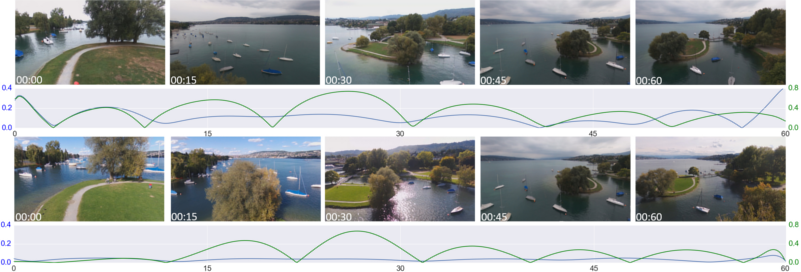}
    \caption{Qualitative comparison of video frames as well as \textcolor{blue}{jerk} (in $\frac{m}{s^3}$) and \textcolor{darkgreen}{angular jerk} (in $\frac{\circ}{s^3}$) profiles of two trajectories generated with \protect\cite{Gebhardt:2016:Airways} (top row) and our method (bottom row).}
	\label{fig:comparison}
\end{figure*}

\section{Implementation}
% Although the system model is linear, the variable discretization step and the moving reference points causes the optimization problem to become non-linear.
We implemented the above optimization problem with MATLAB and solve it with the FORCES Pro software \cite{ForcesPro:2017} which generates fast solver code, exploiting the special structure in the non-linear program.
We set the horizon length of our problem to be $N = 60$.
%in accordance with the task specification of the user study and the default trajectory length of the trajectories of the online survey.
The solver requires a continuous path parametrization. To attain a description of the reference spline across the piecewise sections of the PCHIP spline, we need to locally approximate it. Therefore, we implemented an iterative programming scheme able to generate trajectories at interactive rates. For further details on the IP-scheme and the empirically derived weights of the optimization problem, we refer the interested reader to Appendix \ref{sec:implementation_details}.

\section{Technical Evaluation}
\label{sec:technical_eval}
To evaluate the efficacy of our method in creating smooth camera motion even on problematic (for hard-constrained methods) inputs, we designed a challenging shot and generated two trajectories, one with \cite{Gebhardt:2016:Airways} and the other one with our method.

\begin{figure}[b]
	\centering
	\includegraphics[width=1.0\linewidth]{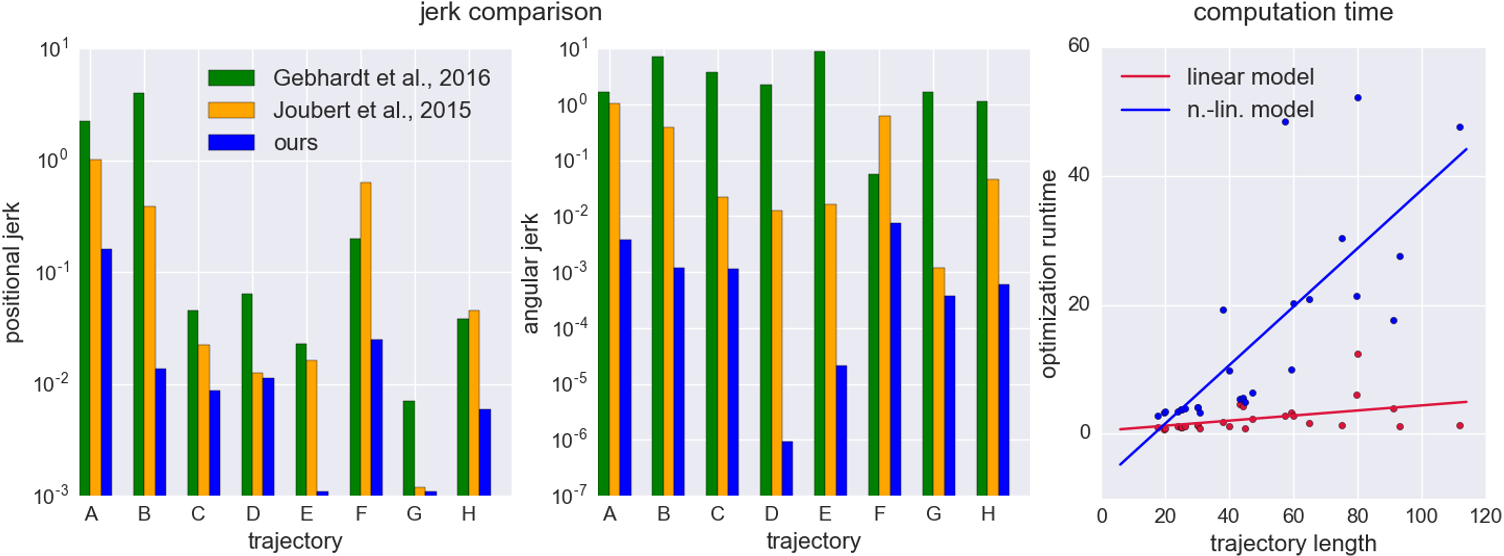}
    \caption{Left: comparing avg. squared jerk (in $\frac{m^2}{s^3}$) and angular jerk (in $\frac{\circ^2}{s^3}$) per horizon stage of different trajectories for our method, \protect\cite{Gebhardt:2016:Airways} and \protect\cite{Joubert2015} (note that latter uses a different model). Right: our method's optimization runtime for different trajectories is plotted against their temporal length (both in $sec$). We differentiate between using a linear and a non-linear quadrotor model for trajectory generation.}
	\label{fig:tech_eval}
\end{figure}

\figref{fig:posPlot} plots the resulting positions in $x,y,z$ and the corresponding camera angles.
Our method adjusts the timing of the original keyframes ($k_2$-$k_4$) to attain smoother motion over time. This is visible when comparing the x-dimension of ours and \cite{Gebhardt:2016:Airways}. The need to trade-off timing and spatial location is illustrated by the orientation plot (\figref{fig:posPlot}, bottom). The keyframes $k_2,k_3$ have been moved very close to each other which would cause excessive yaw velocities since the quadrotor would need to perform a 180\textdegree~turn.
Since our method trades-off the positional fit it generates smooth motion also for the camera orientation.

We also conducted a qualitative comparison by recording different videos with the same consumer grade drone.
The quadrotor followed trajectories generated with our method and with \cite{Gebhardt:2016:Airways} using the same input.
\figref{fig:comparison} shows resulting video frames and jerk profiles (also see [\href{https://www.youtube.com/embed/tfcnLkJpRtc?start=171}{video}]). 
Although the timing of keyframes was improved for smoothness, our method still generates trajectories with lower magnitudes of positional jerk and less variation in angular jerk.

To assess quantitatively that our method generates smoother camera motion, we compare the averaged squared jerk per horizon stage of user-designed trajectories generated with our method, with \cite{Joubert2015} and with \cite{Gebhardt:2016:Airways}. \figref{fig:tech_eval} shows lower jerk and angular jerk values for our optimization scheme compared to both baseline methods, across all trajectories.
% Our method produced an average value of 0.03$\frac{m^2}{s^3}$, SD = 0.05$\frac{m^2}{s^3}$ with an angular jerk of 0.002$\frac{\circ^2}{s^3}$, SD = 0.003$\frac{\circ^2}{s^3}$.This is lower than the jerk (0.83$\frac{m^2}{s^3}$, SD = 1.50$\frac{m^2}{s^3}$) and angular jerk (3.44$\frac{\circ}{s^3}$, SD = 4.58$\frac{\circ}{s^3}$) of the original data.

Finally, we evaluate the optimization runtime of our method. Therefore, we generated trajectories from the studies of \cite{Gebhardt:2017:WYFIWYG,Joubert2015} using the approximated linear quadrotor model of Sec. \ref{sec:dynamic_model} and the non-linear model of \cite{Nageli:2017:SIGGRAPH}.
We measured runtime on a standard desktop machine (Intel Core i7 4GHz CPU, Forces Pro NLP-solver). The computation time for the trajectories are shown in \figref{fig:tech_eval}.
In average, it took 2.41 s (SD = 2.50 s) to generate a trajectory with the linear model and 14.79 s (SD = 15.50 s) with the non-linear model.

%!TEX root = ../submission.tex

\section{Perceptual Study}
Our technical evaluation shows that the proposed method generates smoother trajectories. However, it has not been validated that the trajectories generated with our method result in aesthetically more pleasing video.
To this end, we conduct an online survey comparing videos which follow user-specified timings, generated with the methods of \cite{Gebhardt:2016:Airways,Joubert2015}, with videos generated by our method. Therefore, we compare user-designed trajectories from prior work \cite{Gebhardt:2017:WYFIWYG,Joubert2015}.
For each question we take the user-specified keyframes of the original trajectory and generated a time-optimized trajectory of the same temporal duration (via \equref{eq:set_end_time}) using our method. We then render videos for the original and time-optimized trajectory using Google Earth (based on GPS-coordinates and camera angles). The two resulting videos are placed side-by-side, randomly assigned to the left or right, and participants state which video they prefer on a forced alternative choice 5-point Likert scale. The five responses are: "shot on the left side looks much more pleasing", "shot on the left side looks more pleasing", "both the same", "shot on the right side looks more pleasing", and "shot on the right side looks much more pleasing". Each participant had to compare 14 videos.

\subsection{Results}
In total, 424 participants answered the online survey.
Assuming equidistant intervals, we mapped survey responses onto a scale from -2 to 2, where negative values mean that the original, timed video is aesthetically more pleasing, 0 indicates no difference and a positive value indicates a more aesthetically pleasing time-optimized video. 
In order to attain interval data, our samples are build by taking the mean of the Likert-type results of the expert and non-expert designed videos per participant. Visual inspection of residual plots did not reveal any obvious deviations from normality.

\begin{figure}[tbh]
	\centering
	\includegraphics[width=1.0\linewidth]{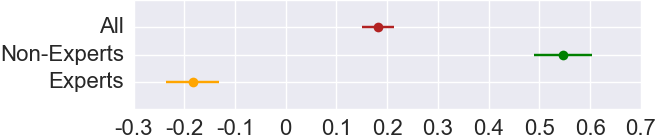}
	\caption{Mean and 95\% confidence interval of the effect of optimization scheme on \emph{all}, \emph{non-expert} designed and \emph{expert} designed videos.}
	\label{fig:online_survey}
\end{figure}

Evaluating \emph{all} responses of the survey, we try to attain a mean which compensates random participant effects. Therefore, we construct a linear mixed model using the participant as random intercept, the video as fixed intercept and introducing a fixed-effect intercept to represent the overall mean. The adjusted mean of the data has a positive value with a high confidence (see Figure \ref{fig:online_survey}). A type III ANOVA showed that there is a significant effect of our method on the aesthetics of videos ($F(1, 856) = 515.4, p < 0.001$).
Unpacking this result further, we distinguish between videos that have been designed by \emph{non-expert} users (data from \cite{Gebhardt:2017:WYFIWYG}) and \emph{expert} users (i.e. cinematographers, data from \cite{Joubert2015}).
Analyzing the results for significance, we perform a one sample t-test on the averaged Likert ratings for \emph{expert}- and \emph{non-expert}-designed videos. The effect of both conditions and their confidence intervals are shown in Figure \ref{fig:online_survey}. While they are significant for both conditions (\emph{expert}: $t(423), p < 0.001$; \emph{non-expert}: $t(423), p < 0.001$), the effect is positive and amplified for \emph{non-expert} designed videos and negative for \emph{expert} designed videos.

% In the case of \emph{non-expert} designed videos the positive effect of our method on aesthetics is amplified, whereas \emph{expert} designed videos are rated higher with timed methods. 
%  \edit{Analyzing the results for significance, we construct a linear mixed model using the participant as random intercept, the video as fixed intercept and introducing a fixed-effect intercept to represent the overall mean. P-values are obtained by performing a generalized linear hypothesis test between \emph{expert}- and \emph{non-expert}-designed videos with the fitted model. Results show a strong significant difference of the effect of conditions on perceived video preference ($EST=0.73, SE=0.03, CI=[0.68, 0.79], z = 25.9, p < 0.001$).}

\subsection{Discussion}
% The perceptual study has shown that videos by \emph{non-experts} are perceived to be more pleasant when generated with our method compared to existing tools. 
The perceptual study provides strong evidence that our method has a positive effect on the aesthetic perception of aerial videos. Furthermore, it has shown that this effect is even stronger for videos by \emph{non-experts}.
This supports our hypothesis that \emph{non-experts} benefit from generating trajectories according to global smoothness as main criteria.
Looking at \emph{expert} created videos the picture is different. These videos were rated as more pleasant when generated with methods which respect user-specified timings. This can be explained by the fact that \emph{experts} explicitly leverage shot timings to create particular compositional effects. Optimizing for global smoothness removes this intention from the result.
% However, the positive mean of \emph{all} responses
However, the significant positive effect of our method on \emph{all} responses
and a larger effect size for the positive effect of \emph{non-expert-} compared to the negative effect of \emph{expert} designed videos indicate that smooth motion is a more important factor for the aesthetic perception of aerial videos than timing.  
This suggests that users, especially experts, could benefit from a problem formulation which allows for soft-constrained instead of hard-constrained timings.
In this way, users could still employ shot timings to create compositional effects, while the optimization scheme generates trajectories trading-off user-specified timings and global smoothness.

Based on these results, we formulate three requirements for quadrotor camera generation schemes:
\begin{inparaenum}[1)]
  \item smoothness should be the primary objective of quadrotor camera trajectory generation,
	\item methods should auto generate or adjust keyframe timings to better support non-experts,
	\item while providing tools for experts to specify soft-constrained timings.
\end{inparaenum}
The proposed method already full-fills requirement 1) and 2). In the next section, we propose how our method can be extended such that \emph{all }requirements are met.

%!TEX root = ../proceedings.tex

\section{Method Extensions}
Recognizing the need to provide both global smoothness and explicit user control over camera timings, we present two method extensions to control camera motion: an approach based on "classic" keyframe timings and a further approach based on velocity profiles.

\subsection{Keyframe Timings}
We augment our objective function with an additional term for soft-constraint keyframe timings. 
The original formulation does not allow for the setting of timing references based on horizon stages: due to the variable horizon we lack a fixed mapping between time and stage. 
To be able to map timings with the spatial reference, we use the $\theta$-parameterization of the reference spline. Reference timings hence need to be specified strictly monotonically increasing in $\theta$. Based on the reference timings and the corresponding $\theta$-values we interpolate a spline through these points, which results in timing reference function $t_d(\theta)$ which can be followed analogously to spatial references by minimizing the cost,
\begin{equation}
	\label{eq:timing_cost}
	c^t(\theta, i, \Delta t) = ||t_d(\theta) - (i * \Delta t)||^2,
\end{equation}
where $i$ is the current stage of the horizon and $\Delta t$ is the current discretization of the model. We add this cost into \eqref{eq:objective_function} and assign a weight to specify its importance $w_t$. By setting the value of $w_t$ to a very large number, quasi hard-constrained keyframes are attainable.

\subsection{Reference Velocities}
The above extension enables mimicry of timing control in prior methods. However, the actual purpose of specifying camera timings in a video is to control or change camera velocity to achieve a desired effect (recall the fly-by example). Since determining the timing of the shot explicitly is difficult, we propose a way for users to directly specify camera velocities.
We extend the formulation of our method to accept reference velocities as input. Again, we use the $\theta$-parameterization to assign velocities to the reference spline $\mathbf{f}_d$. To minimize the difference between the velocity of the quadrotor and the user-specified velocity profile $v_d(\theta)$, we specify the cost,
\begin{equation}
	\label{eq:velocity_cost}
	c^v(\theta, i, \Delta t) = ||v_d(\theta)-\dot{\mathbf{r}}_i^T\mathbf{n}||^2,
\end{equation}
where we project the current velocity of the quadrotor $\dot{\mathbf{r}}_i$ on the normalized tangent vector of the positional reference function $\mathbf{n}$. 
We add this cost term and a weight $w_v$ to \eqref{eq:objective_function}. %The velocity spline is again approximated via local quadratic fits.

%!TEX root = ../proceedings.tex

% \begin{figure}[tbh]
% 	\centering
% 	\includegraphics[width=1.0\linewidth]{figures/tool}
% 	\caption{Screenshot of the quadrotor camera tool used in the study.}
% 	\label{fig:tool}
% \end{figure}

\section{User Study}
To understand whether our final method has the potential to improve the usability of quadrotor camera tools, whether soft-constrained timing methods produce videos of similar perceived aesthetics then hard-constrained timing methods and whether experts can benefit from our method, we conduct an additional user study. In this study, we compare different variants of our method with the method of \cite{Gebhardt:2016:Airways} as representative for quadrotor camera optimization schemes which use hard-constrained keyframe timings.

\begin{figure}[tb]
	\centering
	\includegraphics[width=0.9\linewidth]{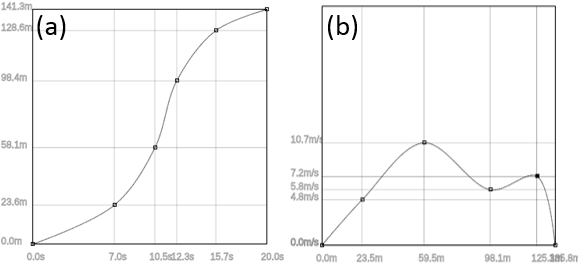}
	\caption{Progress curve (a) and velocity profile (b).}
	\label{fig:toolbar}
\end{figure}

\subsection{Experimental Design}
\paragraph{User Interface:} In our experiment we used the tool of \cite{Gebhardt:2017:WYFIWYG} and extended the UI with a toolbar. This toolbar contains a slider to specify $w_p$ (see \equref{eq:objective_function}) and depending on the condition, a progress curve or a velocity profile. A progress curve allows for the editing of the relative progress on a trajectory over time (see \figref{fig:toolbar}, a). A velocity profile enables editing of the camera speed over the progress on the trajectory (see \figref{fig:toolbar}, b).

\paragraph{Experimental conditions:} We investigate four different conditions:
\begin{inparaenum}[1)]
\item In \emph{timed}, participants work with the optimization method of \cite{Gebhardt:2016:Airways} and a progress curve (see \figref{fig:toolbar}, a).
\item \emph{Soft-timed} uses our optimizer and the progress curve. Participants can decide whether they want to specify keyframe timings (see \equref{eq:timing_cost}) or use the auto-generated timings.
\item In \emph{auto} participants work with our optimization and the keyframe timings it provides. They can choose to fix the end time of a trajectory (see \equref{eq:set_end_time}).
\item  \emph{Velocity} uses our method and a velocity profile (see \figref{fig:toolbar}, b). Participants can decide whether they want to specify camera velocity (see \equref{eq:timing_cost}) or use the auto-generated speed.
\end{inparaenum}

\paragraph{Tasks:} The study comprises two tasks:
\begin{inparaenum}[1)]
	\item Participants were asked to design a free-form video of a building in a virtual environment (T1).  We asked participants to keep the spatial trajectory as similar as possible across conditions whereas the dynamics of camera motion were allowed to differ. 
%     In this way, differences in participants' perception of conditions can can be better attributed towards their different timing abilities. 
They performed the task in the conditions \emph{timed}, \emph{soft-timed} and \emph{auto}.
	\item Participants were asked to faithfully reproduce an aerial video shot with varying camera velocity (T2). Participants should try to reproduce camera path and dynamics of the reference video. This task was performed with the conditions \emph{timed}, \emph{soft-timed} and \emph{velocity} to investigate the level of control over timing afforded by the different conditions.
\end{inparaenum}
We use a within-subjects design and counterbalance order of conditions within a task to compensate for learning effects.

\paragraph{Procedure:} Participants were introduced to the system and the four conditions and were given time to practice using the tool in a tutorial taks. Participants then solved T1 and T2 in the respective conditions. Tasks were completed when participants reported to be satisfied with the designed shot (T1) or the similarity to the reference (T2).
For each task and condition participants completed the NASA-TLX and a questionnaire on levels of satisfaction with the result.
Finally, a short exit interview was conducted.
A session took on average approximately 70 min (introduction $\approx$ 9 min, tutorial $\approx$ 7 min, T1 $\approx$ 22 min, T2 $\approx$ 23 min).

\paragraph{Participants:} We recruited 12 participants (5 female, 7 male). We purposely included 3 experts: an avid hobby quadrotor videographer, a professional videographer, experimenting with quadrotor videography in his free-time, and a professional quadrotor videographer. The remaining participants reported no experience in aerial or normal photo- or videography.

\begin{figure*}[tbh]
	\centering
	\includegraphics[width=1.0\linewidth]{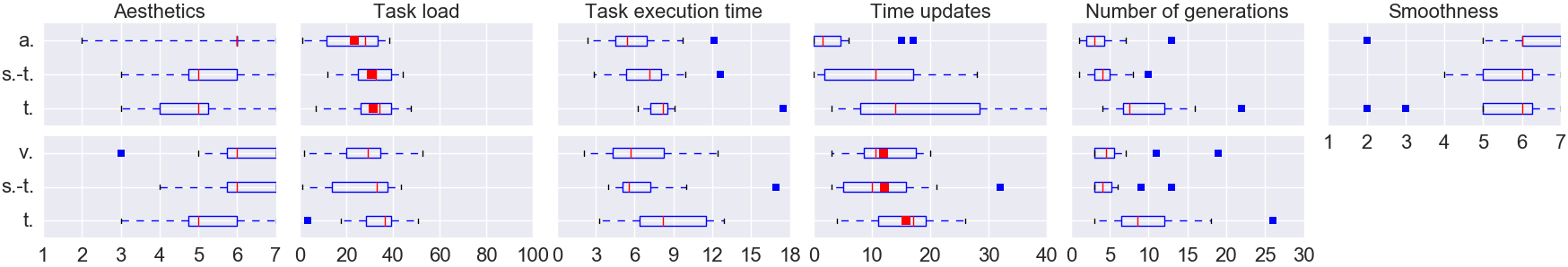}
	\caption{Boxplots for the results of the user study (T1 on upper row, T2 on lower row). The investigated conditions are \emph{auto} (a.), \emph{velocity} (v.), \emph{soft-timed} (s.-t.), and \emph{timed} (t.). Task execution time is depicted in minutes. Aesthetics, smoothness and task load are shown in the respective scales of the questionnaire items. Time updates and number of generations are counts. In case the data of a measure is normally distributed the mean is displayed (red box).}
	\label{fig:boxplots}
\end{figure*}

\subsection{Results}
We analyze the effect of the conditions on the usability of the tool and the aesthetics of the resulting videos. For significance testing, we ran a one-way ANOVA if the normality assumption holds and a Kruskal-Wallis test when it is violated. Analyzing the data of experts and non-experts separately, we found no significant differences in results and thus will not differentiate between them in this section. 

\paragraph{Usability}
To asses the effect of our method on the usability of the tool, we asked participants to fill out NASA-TLX and collected interaction logs (e.g. task execution time). In T1, \emph{auto} has the lowest median in terms of task load, followed by \emph{soft-timed} and \emph{timed} (see \figref{fig:boxplots}).
This ranking remains the same for all interaction logging measures of T1 (see task execution time (TET), time updates and number of generations).
Although there is no significant effect of conditions in T1 on task load ($F(2, 33) = 1.78, p = 0.18$), the other measures do differ significantly (task execution time: $H(2) = 7.38, p < 0.03$; time updates: $H(2) = 10.45, p < 0.01$; number of generations: $H(2) = 13.93, p < 0.01$). 
Pairwise comparison indicates that for TET and time updates \emph{auto} is significantly different to \emph{timed} (TET: $p<0.01$; time updates: $p<0.01$).
For number of generations, \emph{auto} and \emph{soft-timed} significantly differ to \emph{timed} (\emph{auto}-\emph{timed}: $p<0.01$, \emph{soft-timed}-\emph{timed}: $p<0.05$). Other differences are not significant.
\emph{Auto} automatically generates timings and thereby camera velocities. This explains the condition's first rank in terms of task load and interaction logs as it simplifies the task drastically.
For T2, \emph{velocity} and \emph{soft-timed} yield a lower task load compared to \emph{timed}, indicating a slight advantage of our method in terms of usability (differences are not significant: $H(2) = 1.73, p = 0.42$). This ranking is confirmed by interaction logs where \emph{soft-timed} and \emph{velocity} perform similar and are followed by \emph{timed}. The number of generations between conditions differs statistically significantly ($H(2) = 8.04, p < 0.02$).   
A pairwise comparison indicates that \emph{velocity} and \emph{soft-timed} significantly differ from \emph{timed} (\emph{velocity}-\emph{timed}: $p<0.03$; \emph{soft-timed}-\emph{timed}: $p<0.01$). Other differences are not significant (TET: $H(2) = 2.49, p = 0.29$; time updates: $F(2, 33) = 1.16, p =0.33$).
% \CG{The lower number of generations for the conditions \emph{soft-timed}, \emph{auto} and \emph{velocity}) compared to \emph{timed} can also be explained by the fact that our method can generate quadrotor trajectories even if the underlying user input would not yield a feasible result. Due to the soft-constrained timings ts allows our optimization to re-time keyframes such that the output gets feasible.[LEAVE HERE?]}    

\paragraph{Aesthetics}
We are also interested in participants' perceived difference in aesthetics of the generated videos. We asked participants in both tasks to assess the visual quality of the video they designed on a scale ranging from 1 (not at all pleasing) to 7 (very pleasing, see \figref{fig:boxplots}). 
Although differences are not significant (aesthetics in T1: $H(2) = 4.694, p = 0.096$; aesthetics in T2: $H(2) = 3.589, p = 0.166$),
the variants of our method are perceived to produce aesthetically more pleasing videos in both tasks. For T1, we also asked users to rate the smoothness of videos on a scale from 1 (non-smooth) to 7 (very smooth). \figref{fig:boxplots} summarizes the results which do not differ significantly between conditions ($H(2) = 1.828, p = 0.401$).
%Interestingly, the results on aesthetics and smoothness correlate between both, confirming the importance of smoothness for aesthetically pleasing video.

\subsection{Discussion}
Despite the small sample size of our experiment, the results indicate a positive effect of our method on both, the perceived aesthetics of results and the usability of the tool. 
\emph{Auto} caused the lowest task load among conditions and participants where satisfied with the generated results. 
%A reasonable high mean and a tight CI indicate that even experts appreciated its generated results.
Although \emph{soft-timed} and \emph{timed} allow to specify the timing of a shot in the same manor (using the progress-curve or the timeline), \emph{soft-timed} performed better than \emph{timed} in terms of task load (T2) and aesthetics (T1 and T2). We think that this preference can be explained by two factors. First, participants in \emph{soft-timed} generally used a workflow in which they initially take generated timings and then adjust keyframe times to create a desired visual. This workflow was implemented by experts but also by non-experts (if they used keyframe timings at all). Second, in \emph{soft-timed} keyframe timings are specified as soft-constraints, allowing the optimizer to trade-off the temporal fit for a smoother trajectory. This makes \emph{soft-timed} more forgiving than \emph{timed} wrt to the space-time-ratio in-between keyframes, reducing adjustments participants had to do in order to solve a task in this condition (see time/velocity updates and no. of generations in \figref{fig:boxplots}). In addition, soft-constrained timings allow the optimizer to still generate feasible trajectories even if the underlying user input would not yield a feasible result    

The preference for soft-constrained keyframe timings is also an indication for our general assumption that timing control is not used to precisely specify the time a camera should be at a certain position. Instead users employ timing to specify the velocity of the camera along the path. This is also suggested by looking at the results of the \emph{velocity} condition. In T2, it performed similar to \emph{soft-timed} and better than \emph{timed} for task load and aesthetics, indicating that specifying camera dynamics via a velocity profile is an intuitive alternative for providing keyframe timings.

%!TEX root = ../proceedings.tex

\section{Conclusion}
In this paper, we addressed the dichotomy of smoothness and timing control in current quadrotor camera tools. According to design requirements in literature \cite{Joubert2015} their optimizers incorporate keyframes timing as hard constraints, providing precise timing control. A recent study \cite{Gebhardt:2017:WYFIWYG} has shown that this causes users to struggle when specifying smooth camera motion over an entire trajectory. The current optimization formulations needs to have matching distances between the 5 dimensions of a keyframe (position, yaw and pitch of camera angle) with its temporal position. This poses a particular hard interaction problem for users, especially novices.
In this paper, we propose a method which generates smooth quadrotor camera trajectories by taking keyframes only specified in space and optimizing their timings. We formulated this non-linear problem as a variable horizon trajectory optimization scheme which is capable of temporally optimizing positional references.

In a large-scale online survey we compared videos generated with our method to videos generated with \cite{Gebhardt:2016:Airways} and \cite{Joubert2015}. The results indicate a general preference for videos generated according to a global smoothness objective, but also highlight that videos of experts are aesthetically more pleasing when provided timing control.
Based on these insights, we extend our method such that users can specify keyframe timings as soft-constraints but still globally smooth trajectories are attained. In addition, we allow users to specify camera reference velocities set as soft-constraints in the optimization. %problem.

We test the efficacy and usability of our optimization in a comparative user study (baseline is \cite{Gebhardt:2016:Airways}).
The results indicate that our method positively effects the usability of quadrotor camera tools and improves the visual quality of video shots for experts \emph{and} non-experts. Both benefit from using an optimized timing initially and having the possibility of fine-tuning it according to their intention. In addition, the user study revealed that timing control does not need to be precise but is rather used to control camera velocity in order to create a desired compositional effect.

%An interesting direction of future work is to learn the parameters of the proposed optimization scheme.
%Given the scene of a shot, we want to learn the weights of the objective function, further automating the creation of aesthetically pleasing footage, such that it is even easier for users to create videos which follow their compositional intention.

%\section{Acknowledgements}

\begin{acks}
We thank Yi Hao Ng for his work in the exploratory phase of the project, Chat Wacharamanotham for helping with the statistical analysis of the perceptual study and Velko Vechev for providing the video voice-over. We are also grateful for the valuable feedback of Tobias N{\"a}geli on problem formulation and implementation.
This work was funded in parts by the \grantsponsor{SNSF2018}{Swiss National Science Foundation}{http://www.snf.ch/en/Pages/default.aspx} (~\grantnum{SNSF2018}{UFO 200021L\_153644}).
\end{acks}

\newpage
\bibliographystyle{ACM-Reference-Format}
\bibliography{submission}

\appendix
\section{Notation}
\label{sec:notation}
For completeness and reproducibility of our method we provide a summary of the notation used in the paper in \tabref{tab:notation}.

\begin{table}[h]
\centering
\begin{tabular}{l | l}
\textbf{Symbol} & \textbf{Description} \\\hline
$\mathbf{r}$, $\dot{\mathbf{r}}$, $\ddot{\mathbf{r}}$, $\dddot{\mathbf{r}}$ & Quadrotor position, velocity, acceleration and jerk \\
$\psi_q$, $\dot{\psi_q}$, $\ddot{\psi_q}$, $\dddot{\psi_q}$ & Quad. yaw and angular velocity/acceleration/jerk \\
$\psi_g$, $\dot{\psi_g}$, $\ddot{\psi_g}$, $\dddot{\psi_g}$ & Gimbal yaw and angular velocity/acceleration/jerk \\
$\phi_g$, $\dot{\phi_g}$, $\ddot{\phi_g}$, $\dddot{\phi_g}$ & Gimbal pitch and angular velocity/acceleration/jerk \\
$\mathbf{x}$, $\mathbf{u}$ & Quadrotor states and inputs \\
$A$, $B$& System matrices of quadrotor \\
$g$ & Gravity \\
$T$ & Trajectory end time \\
$N$ & Horizon length \\
$\theta$& Progress parameter  \\
$\mathbf{f}_d(\theta)$ & Reference spline ($\mathbb{R}^{5}$) \\
$\mathbf{r}_d(\theta)$ & Positional reference ($\mathbb{R}^{3}$) \\
$\psi_d(\theta)$ & Pitch reference \\
$\phi_d(\theta)$ & Yaw reference \\
$t_d(\theta)$ & Time reference \\
$v_d(\theta)$ & Velocity reference \\
$\mathbf{\Theta}$, $v$& Progress state and input \\
$C$, $D$ & System matrices of progress\\
$\hat{\epsilon}^l$, $\hat{\epsilon}^c$& Approximate lag and contour error
\end{tabular}
\caption{Summary of notation used in the body of the paper}\label{tab:notation}
\end{table}

\section{Implementation Details}
\label{sec:implementation_details}
In this section, we provide details on the weights we use in the objective function, the iterative programming scheme we implemented to attain a continuous path parametrization and its performance.

\subsection{Optimization Weights}
The values for the weights of the objective function we used in the user study and the online survey are listed in \tabref{tab:weights}. 

\begin{table}[tbh]
	\centering
	\setlength{\tabcolsep}{3pt}
	\begin{tabular}[c]{|l||l|l|}
		\hline
        &\multicolumn{2}{|l|}{Value}\\
        \cline{2-3}
		Weight (layed on)&Online survey&User study\\
		\hline
        $w_p$ (position)&1&$[0.1, 10]$ (user-specified)\\
        $Q$ (lag, contour err.)&$diag(2,1)$&$diag(2,1)$\\
		$w_{\psi}$ (yaw)&1&$[0.1, 10]$ (user-specified)\\
		$w_{\phi}$ (pitch)&1&$[0.1, 10]$ (user-specified)\\
  		$w_j$ (jerk)&10&100, 10 (if $w_t, w_v > 0$) \\
		$w_T$ (end-time)&0&1, 0 (if $w_t, w_{end} > 0$)\\
        $w_{end}$ (length in t.)&1&1\\
		$w_t$ (timing)&0&100\\
        $w_v$ (velocity)&0&100\\
        \hline
	\end{tabular}
	\caption{Values for weights used in \equref{eq:objective_function}.}
	\label{tab:weights}
\end{table}

\subsection{Iterative Programming Scheme}
\begin{figure}[tbh]
	\centering
	\includegraphics[width=1.0\linewidth]{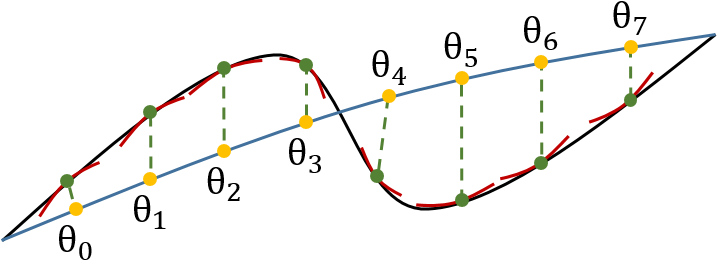}
	\caption{Illustration of the iterative programming scheme, showing the reference path (black) as well as the quadratic approximations of the first (blue) and second (red) iteration with their respective $\theta$-values (yellow, green).}
	\label{fig:it_prog}
\end{figure}

The solver requires a continuous path parametrization. To attain a description of the reference spline even in-between the piecewise sections of the PCHIP-spline, we need to locally approximate it. Therefore, we implement an iterative programming scheme where we compute a quadratic approximation of the reference spline around the $\theta_i$-value of each stage in the horizon. This process is described in \figref{fig:it_prog}. 
In the first iteration of the scheme we initialize all $\theta_i$ to zero and fit the entire reference trajectory (black spline) with a single quadratic approximation (blue spline). By solving the optimization problem of \equref{eq:optimization_problem}, the progression of $\theta$-values will be decided based on the quadratic approximation (yellow dots). For the next iterations, we always take the value of $\theta_i$ from the last run of the solver, project it on the reference spline (green dots) and fit a local quadratic approximation (red splines). Based on these fits the progress of $\theta$-values again is optimized.
We stop the optimization when the difference of the $\theta$-values for all stages in-between iterations is smaller than a pre-defined threshold, $\mathbf{\theta}_{last} - \mathbf{\theta}_{current} < \delta$.

\end{document}